\documentclass[aps,prl,twocolumn,groupedaddress]{revtex4-1}
\usepackage{graphicx}
\usepackage{amsmath}
\usepackage{dcolumn}
\usepackage{bm}

\begin{document}

\preprint{}

\title{Probing the influence of dielectric environment on excitons in monolayer WSe$_2$: Insight from high magnetic fields}

\author{Andreas V. Stier$^1$, Nathan P. Wilson$^2$, Genevieve Clark$^2$, Xiaodong Xu$^{2,3}$, Scott A. Crooker$^{1*}$}

\affiliation{$^1$National High Magnetic Field Laboratory, Los Alamos, New Mexico 87545, USA}
\affiliation{$^2$Department of Physics, University of Washington, Seattle, Washington 98195, USA}
\affiliation{$^3$Department of Materials Science, University of Washington, Seattle, Washington 98195, USA}

\begin{abstract}
Excitons in atomically-thin semiconductors necessarily lie close to a surface, and therefore their properties are expected to be strongly influenced by the surrounding dielectric environment. However, systematic studies exploring this role are challenging, in part because the most readily accessible exciton parameter -- the exciton's optical transition energy -- is largely \textit{un}affected by the surrounding medium. Here we show that the role of the dielectric environment is revealed through its systematic influence on the \textit{size} of the exciton, which can be directly measured via the diamagnetic shift of the exciton transition in high magnetic fields. Using exfoliated WSe$_2$ monolayers affixed to single-mode optical fibers, we tune the surrounding dielectric environment by encapsulating the flakes with different materials, and perform polarized low-temperature magneto-absorption studies to 65~T. The systematic increase of the exciton's size with dielectric screening, and concurrent reduction in binding energy (also inferred from these measurements), is quantitatively compared with leading theoretical models. These results demonstrate how exciton properties can be tuned in future 2D optoelectronic devices.

\end{abstract}
\maketitle

Dielectric screening plays an essential role in semiconductor physics. It modifies the interactions between electronic carriers and therefore strongly impacts not only transport phenomena, but also optoelectronic properties via its influence on both the size and binding energy of bound electron-hole pairs (excitons).  In 3D bulk semiconductors that are characterized by a single dielectric constant $\varepsilon$, exciton radii and binding energies scale simply as $\varepsilon$ and $1/\varepsilon^2$, respectively. For example in  bulk GaAs, the large dielectric constant ($\varepsilon \simeq 13 \varepsilon_0$, where $\varepsilon_0$ is the vacuum permittivity), together with the light electron mass, leads directly to large excitons ($\sim$15~nm radius) with small binding energy ($\sim$5~meV).

In new 2D semiconductors such as monolayer MoS$_2$, WSe$_2$, phosphorene, and germanene -- the recent discoveries of which have sparked tremendous interest \cite{Gomez, MakShan, XuReview, WangReview, Geim, Butler, Tomanek} -- dielectric screening from the semiconductor itself is generally much weaker and is lengthscale-dependent \cite{Keldysh, Cudazzo}. For example, a well-separated electron and hole are essentially \textit{un}screened (because electric field lines connecting the two lie mainly outside the 2D slab), while for electron-hole separations of order the slab thickness many more field lines lie within the slab which partially screens the electrostatic potential. These `non-local dielectric screening' phenomena in 2D semiconductors lead to markedly non-hydrogenic electrostatic potentials $V(r)$ and to excitons with very large binding energies (100s of meV) and correspondingly very small radii ($\sim$1~nm), both in significant contrast to their bulk counterparts \cite{Chei, Rama, Komsa, Berkelbach, Chernikov, Ye, He, Zhu, Ugeda, Zhang, Kyla, WangPRL2015, StierNCOMM, Qiu}.

Crucially, because excitons in 2D semiconductors necessarily reside near a surface, their fundamental properties (size, binding energy, oscillator strength) are expected to be strongly influenced by any \textit{additional} screening from the dielectric environment surrounding the monolayer. These exciton parameters are currently of significant interest, particularly in the monolayer transition-metal dichalcogenide  (TMD) semiconductors \cite{Chei, Rama, Komsa, Berkelbach, Chernikov, Ye, He, Zhu, Ugeda, Zhang, Kyla, WangPRL2015, StierNCOMM, Qiu}. Understanding and quantifying the role of the dielectric environment on 2D semiconductors \cite{Lin, Andersen, Latini} is therefore of critical importance for the design of future optoelectronic devices based on these new materials.

However, studies exploring this role are both scarce and challenging, in part because the most readily accessible property of an exciton -- its optical transition energy -- is largely \textit{un}affected by the surrounding dielectric.  This is because the reduction in exciton binding energy due to the dielectric environment is nearly exactly compensated by an equivalent reduction in the free-particle bandgap, resulting in an unchanged transition energy \cite{Komsa, Kyla, Lin, Latini, Ye}. Current methods for inferring the exciton binding energy have therefore relied on spectroscopy and modeling of excited exciton states \cite{Chernikov, He, Ye, Zhu, WangPRL2015, Qiu}, or a combination of optical measurements and scanning tunneling spectroscopy to determine the free-particle bandgap \cite{Ugeda, Zhang}.

It is desirable, therefore, to identify alternative optical probes of some exciton parameter that is directly impacted by the surrounding dielectric environment, so that its influence can be studied in a quantitative and systematic way.  Fortunately, such a parameter exists: the surrounding dielectric medium is anticipated to directly influence the \textit{size} of the exciton, which in turn can be directly measured via the \textit{diamagnetic shift} of the exciton transition energy \cite{Knox, Miura} in large magnetic fields. It was recently shown that such diamagnetic shifts are indeed observable in monolayer WS$_2$ \cite{StierNCOMM}, but the influence of the surrounding dielectric environment was never varied or studied.

To this end, here we tune the dielectric environment surrounding an archetypal 2D semiconductor (monolayer WSe$_2$), and directly measure its systematic influence on the exciton size via the diamagnetic shift of the A exciton's optical transition energy.  We perform polarized magneto-absorption spectroscopy of exfoliated monolayer WSe$_2$ flakes at low temperature (4~K) and in high magnetic fields to 65~T, and quantify the increasing radius of the A exciton (from $\sim$1.2 to 1.6 nm) as the average dielectric constant of the surrounding environment is increased from 1.55$\varepsilon_0$ to 3.30$\varepsilon_0$. Within the context of a popular theoretical model of nonlocal dielectric screening in these 2D materials (the so-called `Keldysh model'), we also quantify the systematic reduction of the exciton's binding energy (from $\sim$480 to 220~meV). Both of these fundamental exciton parameters are compared with predictions from this widely-used model. While the overall trends are captured quite well, we find that the Keldysh model somewhat underestimates the strong dependence of the measured exciton parameters on environmental screening. These studies therefore provide a route towards predicting the impact of dielectric screening on exciton physics in 2D semiconductors.

Figures 1a-c depict the experimental approach that we developed to perform absorption spectroscopy of monolayer flakes of exfoliated WSe$_2$ at cryogenic temperatures and in the extreme environment of pulsed magnetic fields. The choice of exfoliated WSe$_2$ was motivated by its excellent optical characteristics and large spin-orbit splitting between A and B excitons. An essential requirement in these experiments is to ensure that the light path remains stable and fixed relative to the small monolayer flake during the cooldown of the sample and in the presence of mechanical vibrations that are ubiquitous in pulsed magnets. To accomplish this, we utilized a modified dry transfer process to accurately position and affix an exfoliated flake of monolayer WSe$_2$ over the 3.5 $\mu$m diameter core of a single-mode silica optical fiber. To tune the surrounding dielectric environment, the flake is encapsulated by additional material such as a transparent polymer or (as depicted in Fig. 1) by multilayer hexagonal boron nitride (hBN). The fiber/flake assembly is mechanically very robust, and ensures that light directed through the fiber passes \textsl{only} through the monolayer crystal and does not move with respect to the crystal. Importantly, we note that this new approach can be applied more generally to a broad range of thin materials. Figure 1d shows the room temperature photoluminescence (PL) spectrum of the monolayer WSe$_2$ flake, acquired by pumping with 532~nm light through the optical fiber.

The fiber/flake assembly is mounted on a custom probe and loaded into the cryogenic bore of a 65~T capacitor-driven pulsed magnet at the National High Magnetic Field Laboratory in Los Alamos.  White light from a xenon lamp is coupled into the core of the single-mode fiber, and is transmitted through the monolayer WSe$_2$ flake.  A thin-film circular polarizer after the sample selects only right-circularly polarized light. The transmitted light is then directed back into a 600~$\mu$m diameter collection fiber, and is analyzed by a 300~mm spectrometer and a cooled charge-coupled device (CCD) detector.  Full spectra were acquired every 2.2~ms throughout the $\sim$50~ms long magnet pulse. As shown previously \cite{StierNCOMM, StierJVST}, by detecting right-circularly polarized light in positive magnetic fields to $+65$~T we measure the $\sigma^+$-polarized transition of the A exciton in the $K$ valley of WSe$_2$. To access the $\sigma^-$-polarized exciton transitions in the $K'$ valley, we perform the measurements in the reversed (negative) magnetic field direction (to $-65$~T), which are equivalent by time-reversal symmetry.

\begin{figure}[tbp]
\center
\includegraphics[width=.45\textwidth]{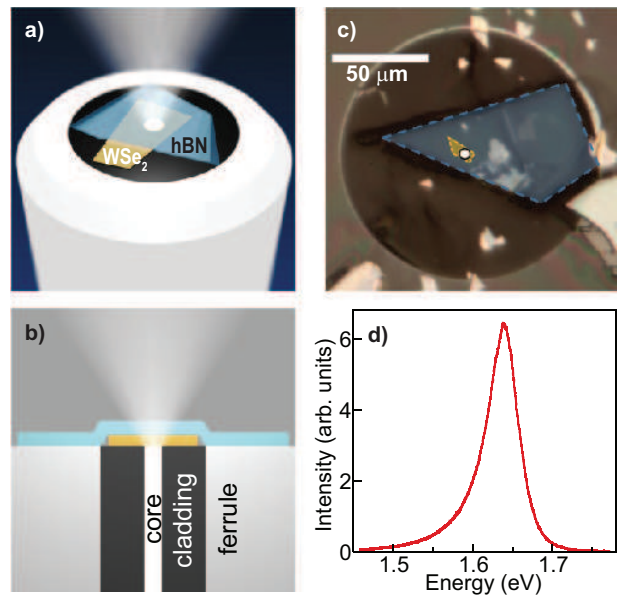}
\caption{(a,b) Experimental schematic: a single exfoliated crystal of monolayer WSe$_2$ is transferred and positioned over the 3.5~$\mu$m diameter silica core of a single-mode optical fiber (not drawn to scale). To tune the dielectric environment, the WSe$_2$ flake is then encapsulated with either hBN (as depicted) or other material such as a transparent polymer. The resulting assembly is physically robust and ensures that light passes only through the monolayer flake and does not move with respect to the flake, even in the cryogenic bore of a 65~T pulsed magnet. (c) Color-enhanced image of the sample/fiber assembly (top view), showing the WSe$_2$ flake (yellow) that is positioned over the fiber core, the hBN overlayer (blue), the 125 $\mu$m diameter fiber cladding (large dark circle), and the ceramic ferrule into which the fiber is epoxied. (d) Room temperature photoluminescence spectrum of the monolayer WSe$_2$ flake, acquired by exciting through the fiber, confirms the monolayer thickness of the flake and the correct position of the flake over the fiber core.} \label{fig1}
\end{figure}

To measure the influence of dielectric screening on the exciton properties, we studied three monolayer WSe$_2$ samples prepared with different surrounding dielectric environments. We define $\varepsilon_b$ and $\varepsilon_t$ as the relative dielectric constants of the bottom substrate and the top encapsulation overlayer, respectively (in units of $\varepsilon_0$, the vacuum permittivity). The first sample, depicted in Figure 1, is the most highly screened: the exfoliated monolayer WSe$_2$ flake sits on the silica fiber ($\varepsilon_b=2.1$) and is encapsulated by a thick 10~nm layer of hBN ($\varepsilon_t=4.5$ \cite{Geick}). The second sample is also on a silica fiber but is encapsulated with a lower-dielectric transparent polymer (polybisphenol carbonate; $\varepsilon_t=2.4$). The third `control' sample is a large-area film of monolayer WSe$_2$ grown by chemical vapor deposition on a SiO$_2$/Si substrate ($\varepsilon_b=2.1$). It is not encapsulated ($\varepsilon_t=1$) and is therefore the most weakly screened.

Note that we use the high frequency (infrared) values for the various dielectric constants, rather than static values (\textit{e.g.}, $\varepsilon_{\rm{silica}}$=2.1, rather than 3.9 in the static limit). This is because the characteristic frequency scale at which a dielectric responds to an exciton is given approximately by the exciton binding energy \cite{Bechstedt}, which is very large -- hundreds of meV -- in monolayer WSe$_2$. This frequency scale exceeds the typical optical phonon frequencies of the surrounding dielectrics (\textit{i.e.}, the lattice cannot respond), therefore optical/infrared values for $\varepsilon$ are more appropriate when considering excitons in monolayer TMD semiconductors. (Note this is in contrast to conventional semiconductors like GaAs, where exciton binding energies are much smaller and static dielectric constants typically suffice.)

Figure 2a shows the low-temperature polarized transmission spectra through the first (hBN encapsulated) sample at $+65$~T, 0~T and $-65$~T. The A exciton absorption resonance of monolayer WSe$_2$ appears as a well-defined minimum in the spectra at 1.73~eV. Magnetic fields break time-reversal symmetry and therefore split the degeneracy between the nominally time-reversed pairs of exciton optical transitions in the $K$ and $K'$ valley: this is the valley Zeeman effect \cite{Aivazian, Srivastava, Wang}.  As the data clearly show, the A exciton absorption resonance in positive fields (hereinafter called $E^+$) shifts to lower energy, while the absorption resonance in negative magnetic fields ($E^-$) shifts to higher energy. Figure 2b shows the $E^+$ and $E^-$ energies versus magnetic field, revealing the valley Zeeman splitting between A excitons in the $K$ and $K'$ valley. The \emph{difference} between the two resonances, $E^+ - E^-$, reveals a valley Zeeman splitting that increases linearly with magnetic field up to 65~T at a rate of $-235 \pm 5$~$\mu$eV/T, indicating an effective valley g-factor $g_v=-4.05 \pm 0.10$ for monolayer WSe$_2$.  This value is in good agreement with, and extends previous results from, low-field PL studies of monolayer WSe$_2$ \cite{Srivastava, Wang, MitiogluNL}

\begin{figure}[tbp]
\center
\includegraphics[width=.4\textwidth]{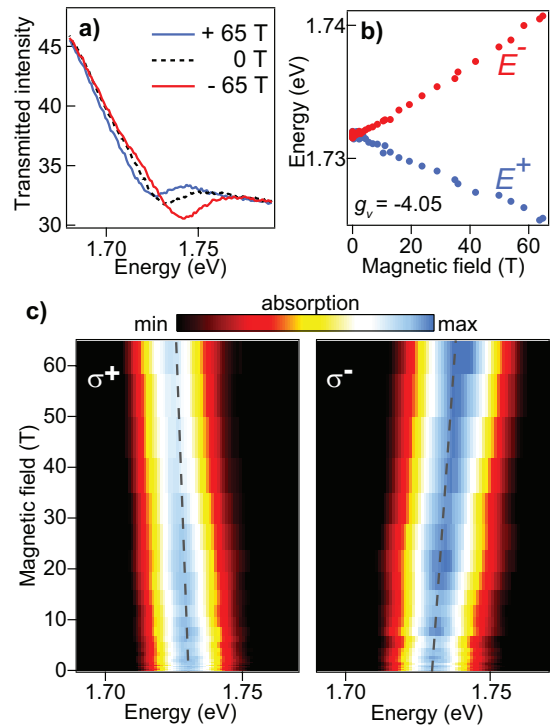}
\caption{(a) Optical transmission spectra at low temperature (4~K) of right-circularly polarized light through the hBN-encapsulated WSe$_2$ flake at $B=0$~T (dashed black curve) and at $B=\pm65$~T (blue, red curves). The valley Zeeman splitting of the A exciton transition is readily apparent. (b) The energy of the A exciton absorption resonance in positive and negative fields ($E^+$ and $E^-$, respectively), which corresponds to the $\sigma^+$ and $\sigma^-$ optical transitions in the $K$ and $K'$ valley of monolayer WSe$_2$. The valley Zeeman splitting ($E^+ - E^-$) increases linearly with field, giving a valley \emph{g}-factor $g_v=-4.05$. (c) Evolution of the polarized magneto-absorption spectra of this WSe$_2$/hBN sample with applied field to 65~T.}
\label{fig2}
\end{figure}

Most importantly, we use these high-field spectra to directly determine the influence of the surrounding dielectric environment on the physical \emph{size} of the exciton -- and also on the exciton binding energy -- via the small quadratic \textit{diamagnetic shift} of the A exciton resonance. To access the diamagnetic shift we examine the average energy of the field-split exciton, $(E^+ + E^-)/2$, in all three WSe$_2$ samples. Figure 3 shows the central result of this work: As the dielectric screening around the monolayer WSe$_2$ sample is systematically reduced, the diamagnetic shift falls by approximately a factor of two, from 0.32 to 0.18 $\mu$eV/T$^2$.  As discussed and quantified immediately below, this directly reveals a reduction in the A exciton radius and also a corresponding increase in the exciton binding energy. 

In general, the ground-state energy of an electrostatically-bound pair of particles -- the $1s$ exciton in semiconductors -- increases quadratically in an applied magnetic field $B$ \cite{Knox, Miura}. This is the exciton diamagnetic shift $\Delta E_{\rm dia}$, which is expressed as
\begin{equation}
\Delta E_{\rm dia} = \frac{e^2}{8 m_r} \langle r^2 \rangle_{1s} B^2 = \sigma B^2.
\end{equation}

Here, $\sigma$ is the diamagnetic shift coefficient, $m_r$ is the reduced mass of the exciton ($m_r^{-1} = m_e^{-1} + m_h^{-1}$), and $r$ is a radial coordinate in the plane perpendicular to $B$ (here, this is the plane of the 2D monolayer). The term $\langle r^2 \rangle_{1s}$ denotes the expectation value of $r^2$ over the $1s$ exciton wavefunction $\psi_{1s}(r)$; namely, $\langle \psi | r^2 | \psi \rangle$. Equation (1) applies in the so-called `weak-field limit', where the characteristic magnetic energy scales $\Delta E_{\rm dia}$ and $\hbar \omega_c$ (the cyclotron energy) are much less than the exciton binding energy. Due to the huge exciton binding energies in monolayer TMDs, this limit applies even in large 65~T magnetic fields.

Given the exciton's reduced mass $m_r$, the root-mean-square (rms) radius of the $1s$ exciton in the monolayer plane, $r_X$, is therefore given by

\begin{equation}
r_X \equiv \sqrt{\langle r^2 \rangle_{1s}} = \sqrt{8 m_r \sigma}/e.
\end{equation}

This basic result of semiconductor physics is entirely general and is, crucially, independent of the functional form of the electrostatic binding potential $V(r)$ and therefore the shape of $\psi_{1s}(r)$. [Note that $r_X$ is related to the exciton ``Bohr radius'' $a_0$, a notion that applies primarily in bulk semiconductors having conventional Coulomb potentials $V(r) \propto -1/r$, for which case $r_X = \sqrt{2} a_0$].

\begin{figure*}[tbp]
\center
\includegraphics[width=.55\textwidth]{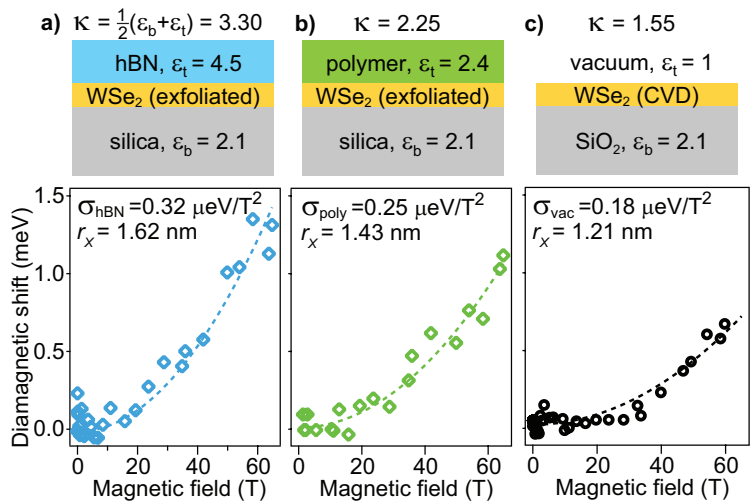}
\caption{The plots show the \emph{average} energy of the field-split exciton transitions, ($E^+$+$E^-$)/2, versus applied magnetic field to 65~T. The average energy increases quadratically with field, revealing the diamagnetic shift ($\sigma$) in monolayer WSe$_2$, from which the exciton radius $r_X$ (and also the exciton binding energy) can be inferred.  The  diamagnetic shift is measured for monolayer WSe$_2$ in the three different dielectric environments depicted by the diagrams: (a) WSe$_2$ on silica and encapsulated by hBN, (b) WSe$_2$ on silica and encapsulated by a lower-dielectric polymer, and (c) a `control' sample of CVD-grown monolayer WSe$_2$ on SiO$_2$/Si (this sample was measured using magneto-reflection spectroscopy techniques developed in Refs. \cite{StierNCOMM, StierJVST}). The systematic decrease of the diamagnetic shift from (a) to (c) reveals the corresponding decrease of the exciton radius (and concurrent increase in exciton binding energy) as the average dielectric constant of the material surrounding the WSe$_2$ flake, $\kappa=(\varepsilon_b + \varepsilon_t)/2$, is reduced. Values of $r_X$ are computed from Eq. 2, using $m_r = 0.18 m_e$ (see text).}
\label{fig3}
\end{figure*}

Notably, Equation (2) does not depend explicitly on the assumption of any dielectric properties: The exciton radius $r_X$ is directly determined from the diamagnetic shift alone. For example, using a reduced exciton mass of $0.18 m_e$ (a value recently obtained from density-functional theory \cite{Kyla}), we therefore find that $1s$ excitons in WSe$_2$ have rms radii $r_X$=1.62 nm, 1.43 nm, and 1.21 nm for the three samples shown in Figure 3 that are overcoated with hBN, polymer, and nothing respectively. The shrinkage of the exciton is due to the reduction of dielectric screening by the surrounding environment.  These studies therefore provide the first explicit measurement of exciton size in monolayer WSe$_2$, and the first systematic study of how this fundamental exciton property is related to changes in the dielectric environment surrounding a monolayer semiconductor.

Having established that the dielectric environment does indeed directly influence the \textit{size} of excitons in monolayer semiconductors, we now seek to quantitatively compare our experimental data with current theoretical models. We therefore adopt a commonly-used framework, described below, to model the non-hydrogenic electrostatic potential $V(r)$ that binds electrons and holes in 2D semiconductors.  Furthermore, using this model, we discuss how diamagnetic shift measurements can also be used to quantify the influence of the surrounding environment on the exciton binding energy itself, another parameter of considerable current interest in monolayer semiconductors \cite{Chei, Rama, Komsa, Berkelbach, Chernikov, Ye, He, Zhu, Ugeda, Zhang, Kyla, WangPRL2015, StierNCOMM, Qiu}.

In contrast to bulk materials, the effective material dielectric constant that is `seen' by a bound electron-hole pair in a 2D material depends strongly on their separation $r$. This `non-local dielectric screening', which has been discussed extensively in recent literature of 2D materials \cite{Cudazzo, Berkelbach, Chernikov, Ye, Latini, Qiu}, strongly modifies the electrostatic potential $V(r)$ between an electron and a hole, and leads to a non-hydrogenic Rydberg series of excited exciton states. Following the original formulation by Keldysh \cite{Keldysh}, we consider our monolayer WSe$_2$ samples as infinitely thin slabs that are bounded below and above by material with relative dielectric constants $\varepsilon_b$ and $\varepsilon_t$, respectively.  In this case, the potential $V(r)$ can be expressed analytically as

\begin{equation}
V(r)=-\frac{e^2}{8 \varepsilon_0 r_0}\left[ H_0 \left(\frac{\kappa r}{r_0} \right) - Y_0 \left(\frac{\kappa r}{r_0}\right) \right].
\end{equation}

Here, $H_0$ and $Y_0$ are the Struve function and Bessel functions of the second kind, respectively. The dielectric properties of the 2D material itself are captured by the characteristic screening length $r_0$, which is related to the 2D polarizability $\chi_{2D}$ of the monolayer material: $r_0 = 2 \pi \chi_{2D}$. Recent calculations suggest that $r_0$ lies in the range from 3-5 nm for the family of monolayer TMD semiconductors \cite{Kyla, Berkelbach}. The role of the dielectric environment is captured by $\kappa = (\varepsilon_b + \varepsilon_t)/2$, the average dielectric constant of the surrounding material. We note that this potential applies only when $\kappa$ is less than the dielectric constant of the 2D material itself \cite{Keldysh, Kyla}, which is the case in our studies. This potential follows the conventional (hydrogenic) $1/\kappa r$ dependence at large electron-hole separations $r \gg r_0$, but diverges only weakly as log($r$) at small separations $r \ll r_0$ due to screening from the 2D dielectric sheet. This form of the potential $V(r)$ is frequently used to model excitons in monolayer materials \cite{Komsa,  Berkelbach, Chernikov, Rana, Varga, Velizhanin}, and has been found to approximate reasonably well the more exact functional form of the nonlocal screening that can be computed using \textit{ab initio} methods \cite{Qiu, Latini}.

\begin{figure*}[tbp]
\center
\includegraphics[width=.65\textwidth]{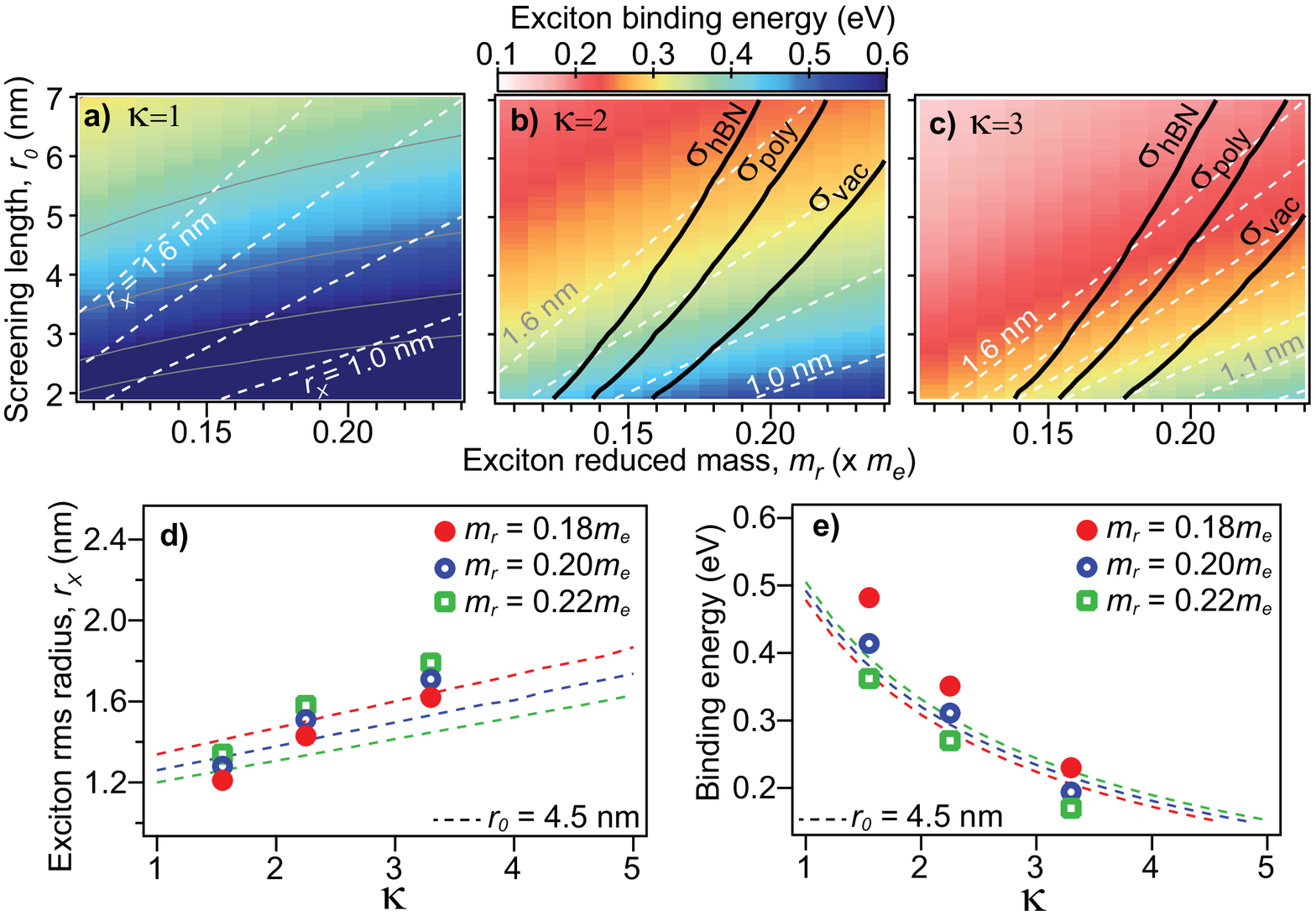}
\caption{(a-c) Color surface plots of the exciton binding energy in a monolayer semiconductor, calculated by solving Schr\"{o}dinger's equation to find the $1s$ exciton wavefunction $\psi_{1s}(r)$, using the screened Keldysh potential $V(r)$ defined in Eq.~(3). Calculations are performed over a range of possible exciton reduced masses $m_r$ and material screening lengths $r_0$. Contours of constant binding energy are marked by thin lines [in panel (a) only], and contours of constant rms exciton radius $r_X$ are indicated by dashed lines. Panels (a), (b), and (c) correspond to increasing $\kappa$, the average dielectric constant of the surrounding environment ($\kappa$=1, 2, and 3 respectively). The thick solid lines are the contours of constant diamagnetic shift corresponding to our experimentally-measured values ($\sigma_{\rm{hBN}}$, $\sigma_{\rm{poly}}$, and $\sigma_{\rm{vac}}$ equal to 0.32, 0.25, and 0.18~$\mu$eV/T$^2$, respectively).  Within this Keldysh model, exciton radius and binding energy are therefore obtained at the intersection of the appropriate $\sigma$ contour with the assumed value of $m_r$. (d) Comparing the measured rms exciton radius $r_X$ (points) as a function of $\kappa$ with expectations from the screened Keldysh model (lines). Calculated values and simulations are shown for three different values of $m_r$. (e) Comparing the exciton binding energy inferred from the diamagnetic shift data (points) with expectations from the screened Keldysh model (lines).  Overall these trends are reproduced rather well, but the Keldysh model underestimates the measured dependence of the exciton radius and binding energy on $\kappa$.} \label{fig4}
\end{figure*}

Using this potential, we then numerically solve the Schr\"{o}dinger equation to calculate the $1s$ exciton wavefunction $\psi_{1s}(r)$, its rms radius $r_X$, its binding energy, and its expected diamagnetic shift $\sigma$ for any input values of reduced mass $m_r$ and screening length $r_0$.  Figure 4a shows the results of such calculations for the case of a suspended 2D layer in vacuum ($\kappa=\varepsilon_{t,b} = 1$).  Different colors (separated by solid contours) indicate the calculated exciton binding energy.  Also shown by dashed lines are contours of constant exciton radius $r_X$. As expected, larger masses $m_r$ and/or smaller screening lengths $r_0$ lead to smaller excitons with larger binding energy.

Figures 4b and 4c show similar surface plots for the cases of $\kappa=2$ and $\kappa=3$, which correspond to increasing screening from the surrounding dielectric environment. As anticipated, for fixed values of $m_r$ and $r_0$, the exciton's radius increases and its binding energy drops as $\kappa$ increases and the surrounding media screens the 2D exciton more effectively. The binding energies and radii that can be extracted from these plots for all the monolayer TMDs are in excellent agreement with the recent calculations of Kyl\"{a}np\"{a}\"{a} \cite{Kyla}, which are also based on the screened Keldysh potential. 

Most importantly, the thick black lines in Figs. 4b,c show the calculated \emph{contours of constant diamagnetic shift} that correspond to the experimentally-measured values that we obtained in our high-field studies (\textit{i.e.}, $\sigma_{\rm{hBN}}$, $\sigma_{\rm{poly}}$, and $\sigma_{\rm{vac}}$ equal to 0.32, 0.25, and 0.18~$\mu$eV/T$^2$, respectively).  The utility of these calculations and surface plots are therefore now evident: Within this model, the binding energy (and also radius) of the A exciton is found at the intersection of the appropriate diamagnetic shift contour with the value of $m_r$ that is used. In this way, experimental measurements of the diamagnetic shift can also be used to significantly constrain estimates of the exciton binding energy.

Using this model, we first compare the measured values of the A exciton radius $r_X$ -- obtained from the diamagnetic shift alone as described above -- with the calculated values of $r_X$ obtained using the screened Keldysh potential in Equation 3. Both are plotted in Fig. 4d as a function of $\kappa$ of the surrounding dielectric environment. To most clearly convey how the inferred values of $r_X$ vary with the exciton reduced mass $m_r$ that is assumed, we plot the measured and simulated $r_X$ for three different values of $m_r$ (0.18, 0.20, and 0.22$m_e$). The simulations use a screening length $r_0=4.5$~nm, following theoretical predictions for monolayer WSe$_2$ \cite{Kyla, Berkelbach}. The overall growth of $r_X$ with increased dielectric screening is captured reasonably well; however, expectations from the Keldysh model somewhat underestimate the stronger dependence of $r_X$ on $\kappa$ that is actually measured.

Similarly, Fig. 4e shows the binding energies $E_B$ of the A exciton that are inferred, using this model, from our diamagnetic shift measurements.  Again, three sets of points are shown, corresponding to different $m_r$. Details of this analysis are also summarized in Table I. As shown, $E_B$ (points) decreases markedly as $\kappa$ is increased. Expectations from the screened Keldysh model (lines) are in reasonable and qualitative agreement (particularly when using $m_r=0.20 m_e$); however, once again the Keldysh model underestimates the stronger dependence of $E_B$ on $\kappa$ that is actually measured. As discussed in previous work \cite{Kyla}, the screened Keldysh model predicts an approximately power-law dependence of $E_B$ on $\kappa$, $E_B = E_B^0/\kappa^\alpha$, with an exponent $\alpha \simeq 0.7$.  Here we find that the data indicate a somewhat stronger power-law dependence with an exponent $\alpha \simeq 0.95$. 

We note that both the model and the data trend towards a binding energy of $\sim$500~meV in the absence of environmental screening ($\kappa=1$; \textit{i.e.}, a monolayer suspended in vacuum), over a range of reasonable exciton masses. This value is in quite decent agreement with some theoretical estimates of exciton binding energy for freestanding WSe$_2$ \cite{Komsa, Berkelbach, Varga}.  Moreover, we can also compare our results for the common experimental case of monolayer WSe$_2$ on a SiO$_2$ substrate (for which $\kappa=1.55$), where recent spectroscopic measurements of excited exciton states \cite{He} inferred a 370~meV binding energy. As shown in Fig. 4e, we obtain a similar value if $m_r=0.22 m_e$ is used, suggesting that the exciton reduced mass in monolayer WSe$_2$ may in fact be $\sim$20\% heavier than its commonly-assumed value of $m_r$=0.17-0.18$m_e$ (which are based on density-functional-theory calculations \cite{Kyla, Berkelbach}).  A heavier exciton mass in monolayer WSe$_2$ would also be consistent with our earlier measurements of monolayer WS$_2$ on SiO$_2$ ($\kappa$=1.55), for which a larger diamagnetic shift was measured \cite{StierNCOMM}.

\begin{table*}
\centering
\begin{tabular}{l c c c c c c}
\hline\hline
Material & $\varepsilon_t$ & $\varepsilon_b$ & $\kappa$ & $\sigma$($\mu$eV/T$^{2}$) & $r_X$(nm) & $E_B$(meV)  \\[0.5ex]
\hline
WSe$_{2}$ (uncapped) & 1.0 & 2.1 & 1.55 & 0.18$\pm$.02 & \begin{tabular}{@{}l@{}}1.21 \\1.34 \end{tabular} & \begin{tabular}{@{}l@{}}482 \\362 \end{tabular} \\[0.5ex]
\hline
WSe$_{2}$ (polymer) & 2.4 & 2.1 & 2.25 & 0.25$\pm$.02 & \begin{tabular}{@{}l@{}}1.43 \\1.58 \end{tabular}  & \begin{tabular}{@{}l@{}}351 \\270 \end{tabular}   \\[0.5ex]
\hline
WSe$_{2}$ (hBN) & 4.5 & 2.1 & 3.30 & 0.32$\pm$.02 & \begin{tabular}{@{}l@{}}1.62 \\1.79 \end{tabular}  & \begin{tabular}{@{}l@{}}221 \\180 \end{tabular}   \\[0.5ex]
\hline
\end{tabular}
\caption{Summary of the dielectric environments ($\varepsilon_t, \varepsilon_b, \kappa$) and measured diamagnetic shifts ($\sigma$) for the three monolayer WSe$_2$ samples used in these studies. Also shown are the $1s$ exciton radii $r_X$ (obtained from $\sigma$ using Eq. 2) and the $1s$ exciton binding energies $E_B$ (inferred from $\sigma$, using the screened Keldysh model as described in the main text). Two values of $r_X$ and $E_B$ are shown for each sample; the upper values are obtained using a reduced exciton mass $m_r=0.18 m_e$, while the lower values are obtained with $m_r = 0.22 m_e$.}
\label{table:sigmas}
\end{table*}

Finally, we wish to emphasize that the exciton wavefunction, its size, and its binding energy are necessarily very sensitive to the exact form of the potential $V(r)$, which in turns depends on the details of the dielectric environment and choice of substrate material \cite{Komsa, Lin}. The screened Keldysh potential of Eq.(3) is only an approximation to the more accurate potential that can be calculated from first principles \cite{Qiu, Latini}. We hope that these experimental results and analysis may inspire additional theoretical and computational studies along these lines, and may also address the question of how the exciton oscillator strength varies with $\kappa$. Nevertheless, the observed trends of increasing diamagnetic shift (indicating increasing exciton size and decreasing binding energy) with increasing dielectric screening from the environment are robust and consistent with expectations. 

In summary, we have studied the influence of the surrounding dielectric environment on the size and binding energy of excitons in an atomically-thin semiconductor (monolayer WSe$_2$) through examination of the exciton's diamagnetic shift in pulsed magnetic fields to 65 T. We quantify for the first time the systematic increase of the exciton size (and reduction of exciton binding energy) with increased environmental screening, which will be important for the design of future optoelectronic devices based on 2D semiconductors. Finally, the experimental techniques we developed (bonding of monolayer semiconductors to single mode optical fibers) should be broadly applicable to a wide variety of new and interesting 2D materials, for studies of fundamental exciton and optical properties. 

We gratefully acknowledge helpful discussions with K. Velizhanin, D. L. Smith, H.-P. Komsa, and H. Dery.  These optical studies were performed at the NHMFL, which is supported by NSF DMR-1157490 and the State of Florida. The work at UW was supported by the U.S. DOE Basic Energy Sciences, Materials Sciences and Engineering Division (DE-SC0008145 and SC0012509).

\end{document}